\documentclass[aip,epsfig,jcp,twocolumn]{revtex4}

\usepackage{float}
\usepackage{epsfig}
\usepackage{graphicx}
\usepackage{subfigure}
\usepackage[dvips]{color}
\begin{document}

\title{Free volume distribution of nearly jammed hard sphere packings}

\author{Moumita Maiti}
\email{maiti@theorie.physik.uni-goettingen.de}

\thanks{Present Address: Institute for theoretical Physics, University of Goettingen, Germany.}

\affiliation{Jawaharlal Nehru Centre for Advanced Scientific Research, Jakkur Campus, Bangalore 560 064, India}
\affiliation{TIFR Centre for Interdisciplinary Sciences, 21 Brundavan Colony, Narsingi, 500075 Hyderabad, India}

\author{Srikanth Sastry}
\email{sastry@jncasr.ac.in}
\affiliation{Jawaharlal Nehru Centre for Advanced Scientific Research, Jakkur Campus, Bangalore 560 064, India}
\affiliation{TIFR Centre for Interdisciplinary Sciences, 21 Brundavan Colony, Narsingi, 500075 Hyderabad, India}


\begin{abstract}
We calculate the free volume distributions of nearly jammed packings of
monodisperse and bidisperse hard sphere configurations. These distributions 
differ qualitatively from those of the fluid, displaying a power law tail at large free volumes, which constitutes a distinct signature of nearly jammed
configurations, persisting for moderate degrees of decompression. We
reproduce and explain the observed distribution by considering the
pair correlation function within the first coordination shell for
jammed hard sphere configurations. We analyze features of the equation
of state near jamming, and discuss the significance of observed asphericities of the free volumes to the equation of state.
\end{abstract}
\maketitle


Disordered, structurally arrested forms of matter arise through a
variety of procesess, such as the glass transition in dense, cold
liquids, gelation, jamming in granular matter, to name a few
examples\cite{debenedetti-book,binder-kob-book,heterogeneity-book,torquato-book}.
The nature of these processes and their interrelation are topics of
considerable current research activity. Much of the interest is
focussed on understanding the manner in which the relaxation dynamics and
transport properties change in the fluid state upon approaching the
arrest transition. Temperature, density and applied stress are
involved as control parameters that determine the transition from an
arrested to a fluid state in different systems, and a unified {\it
  jamming phase diagram} has been discussed as a possible description
of the arrested-to-fluid transition \cite{liu}. Recent theoretical
work, however, draws a distinction between the glass and jamming
transitions (summarized, {\it e. g.} in \cite{parisi-zamponi-review}),
and the exact relationship between these transitions is a topic of
ongoing research. 

Considerable work addressing the nature of the jamming transition has
focussed on packings of hard spheres, dating back to the seminal work
of Bernal and co-workers \cite{bernal,bernalmason}, who identified a
{\it random close packing} (RCP) of spheres with a packing fraction of
$64 \%$, attained as the maximum density for amorphous packings of
equal sized spheres \textcolor{black}{(we express density values in packing fractions throughout)}. Considering the hard sphere fluid as
the initial, thermodynamic equilibrium state from which jammed
configurations are generated, RCP would represent a limit state at
which the pressure would diverge. While such a point would correspond
to a state of structural arrest, many studies suggest that a {\it
  glass} transition precedes jamming, at finite pressure
\cite{speedyjeos,speedyhsgt,krzakala,parisi-zamponi,parisi-zamponi-review}.
Further, in analogy with the glass transition, it has been suggested and demonstrated 
that the jamming transition too does not occur at a unique point but
can occur over a range of densities
\cite{speedyhsgt,krzakala,chaudhuri,sal1}.

In addition to such analyses concerned with thermodynamic and dynamic
(or kinetic) aspects, much attention has been focussed on geometric
features of jammed configurations
\cite{mrj,torquatorev,donev,epitome,silbert,bruic,wyart,sal2,ogarko} (in contrast
to the relatively scant attention to structure in the context of glass
formation). From such a geometric point of view, the proposal of a
unique random close packing has been questioned \cite{mrj}, and it has
been proposed that a more suitable notion is that of maximally random
jammed packings. Although the jamming density itself is not unique,
interestingly, many geometric features remain robust
\cite{chaudhuri}. An important example is the power law singularity in
the pair correlation function near contact, observed in addition to
the delta function at contact \cite{epitome,silbert,donev,wyart,charbonneau}. The
origins and consquences of these unique geometric features are of
considerable current interest.

One of the important ways of characterizing the geometry of hard
sphere packings is the distribution of free volumes
\cite{sastryfreevol,maiti}. The free volume, defined as the volume available
to the centre of a hard sphere while all other
spheres remain fixed, is further directly related to the equation of state (EOS) \cite{speedyhs},
\begin{equation}
{P \over \rho k_B T} = 1 + {\sigma \over 2 D} \left<{s_f \over v_f}\right>,
\label{eq:fveos} 
\end{equation} 
where $P$ is the pressure, $T$ is the temperature, $\sigma$ the diameter of the spheres, $D$ is the spatial dimensionality \textcolor{black}{($D = 3$ for all the results shown in this work.)}, $v_f$ and $s_f$ are respectively the free volume and the surface area of the free volume. An exact algorithm for determining the volume and surface area of cavities in polydisperse sphere packings, developed in \cite{sastryalgo} has been employed \cite{sastryfreevol,maiti} to study free volume distributions in (monodisperse and bidisperse) hard sphere fluids. 
\begin{figure}[H]
\centering
\includegraphics[scale=0.3]{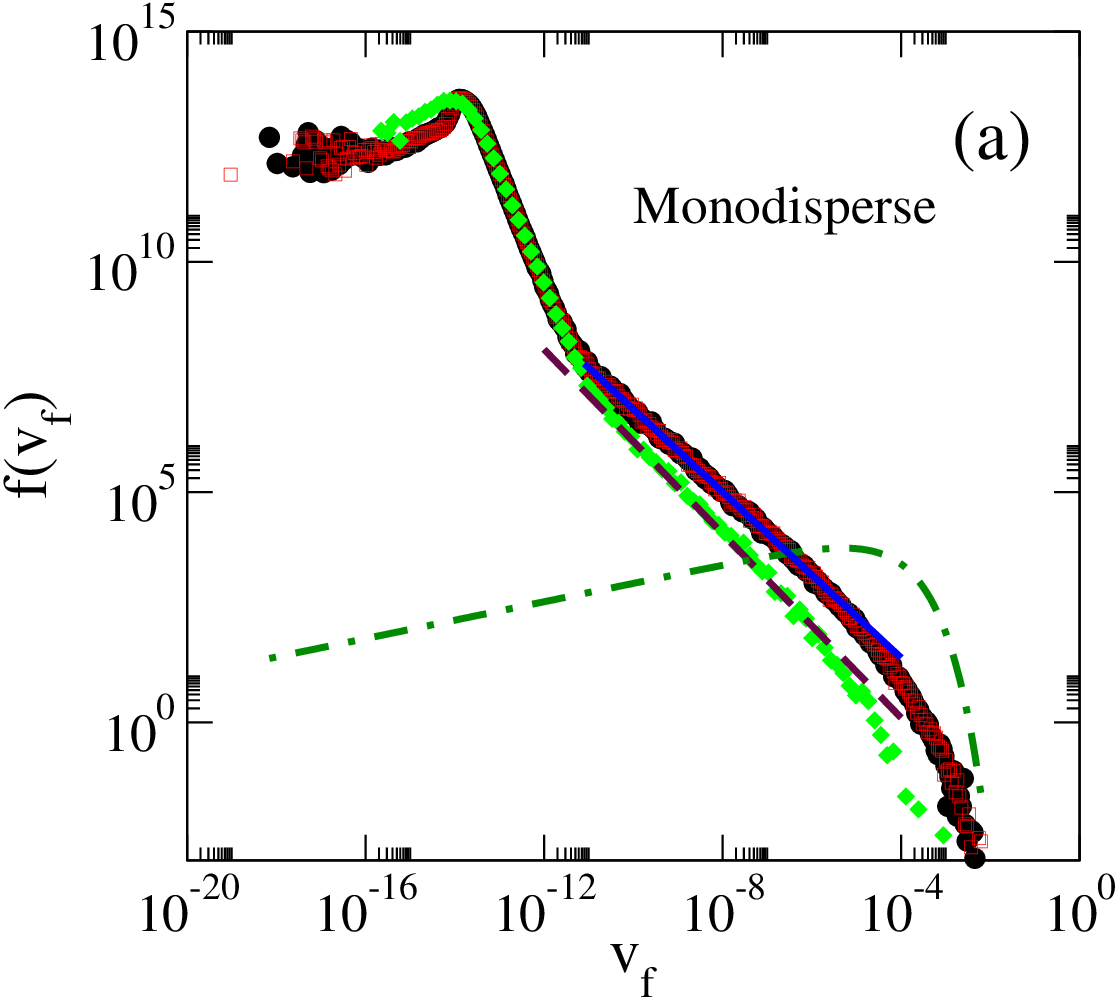}
\includegraphics[scale=0.3]{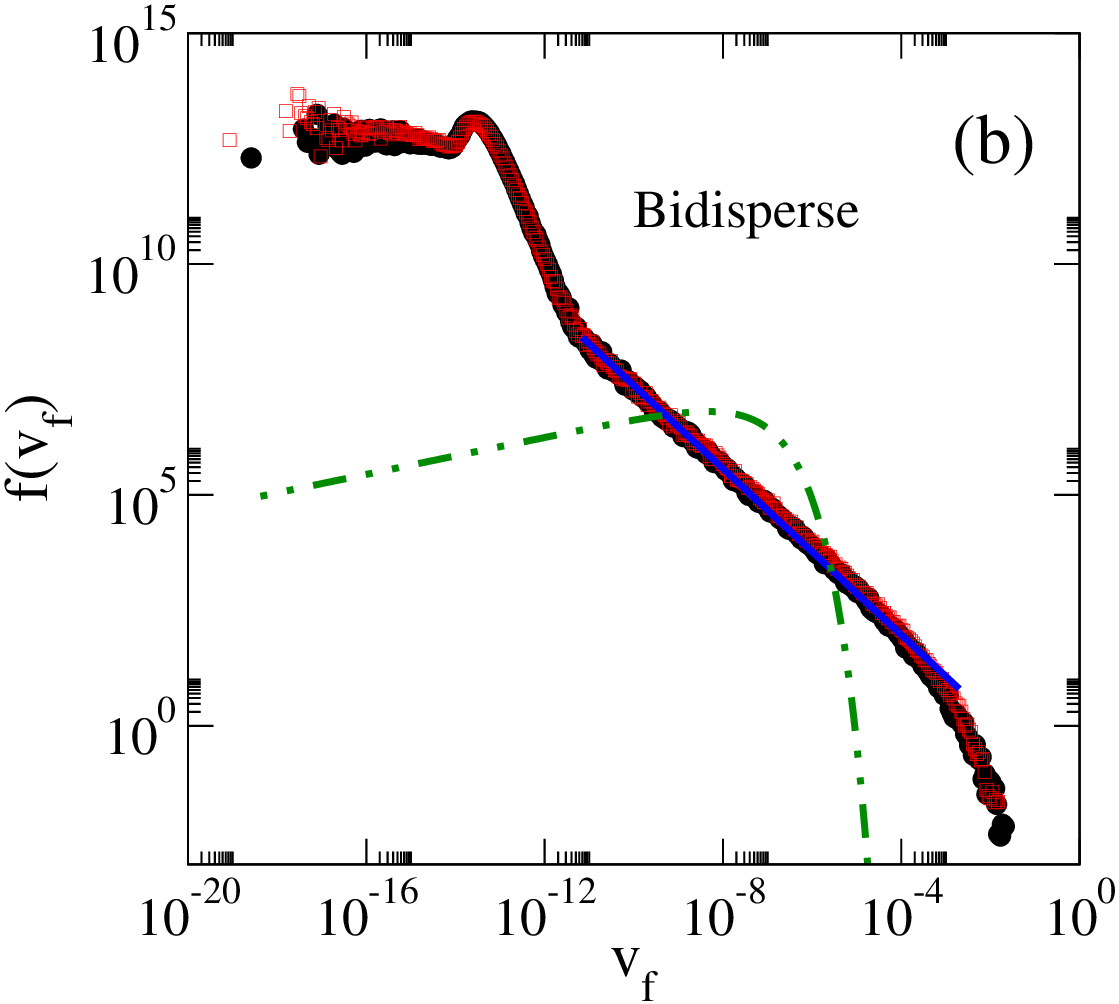}
\includegraphics[scale=0.3]{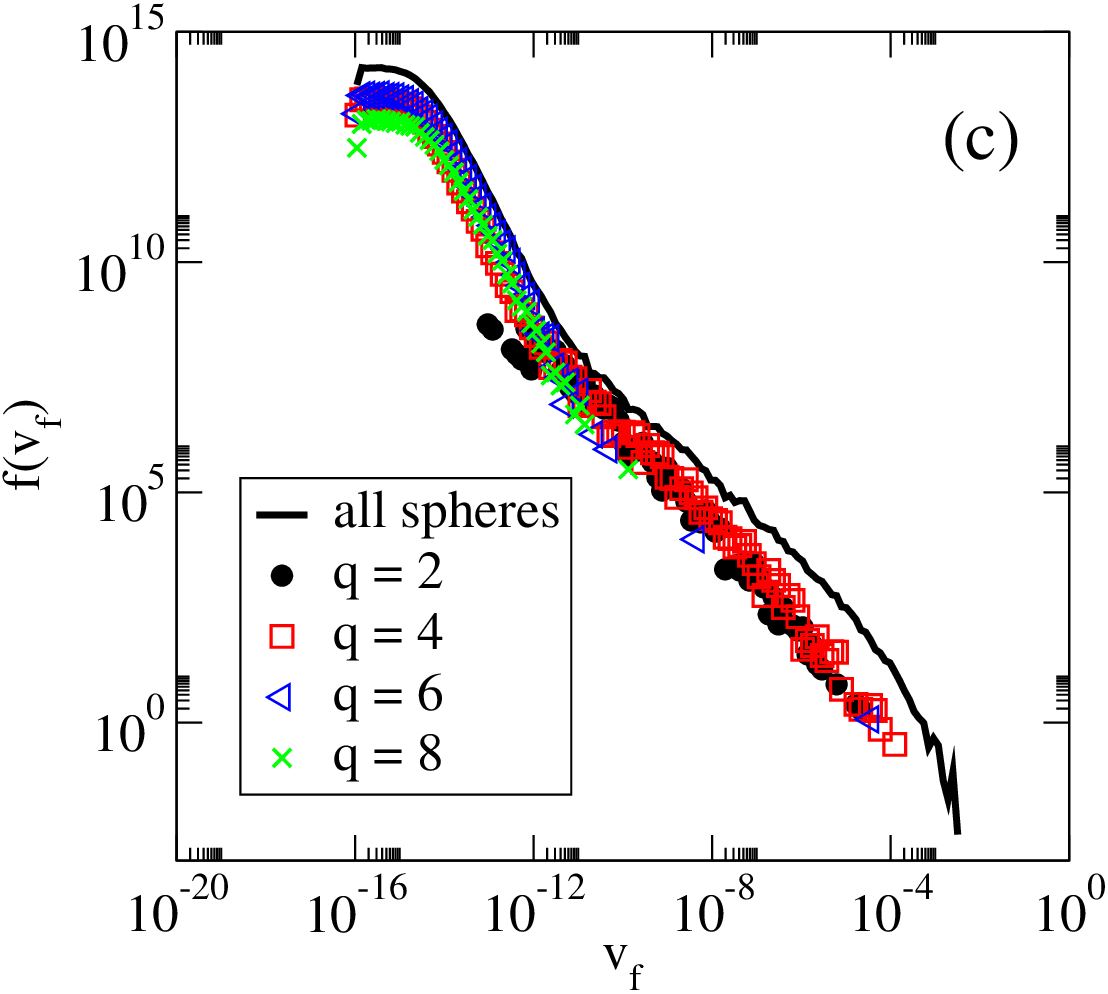}
\caption{Free volume distributions for (a) monodisperse (circles: $\phi_{init} = 0.4487$; squares: $\phi_{init} = 0.533$) and (b) (component 1 of the) bidisperse sphere packings ((circles: $\phi_{init} = 0.363$; squares: $\phi_{init} = 0.5919$) close to the jamming point, for the cases listed in tables I and II. Free volume distributions of both packings have a power law tail of exponent $-0.9$ (solid lines). The distributions are the same regardless of the jamming density. The expected free volume distributions, based on extrapolating \textcolor{black}{(by a polynomial fit to the parameters in Eq. \ref{eq:fv} and extrapolating them to the jamming density)} those for the fluid (dot dashed lines), are shown for comparison. The distribution obtained without considering the free volumes of rattlers is also shown in (a) for one case ($\phi_{init} = 0.4487$) (diamonds), with a power law tail with exponent $-1$) (dashed line). (c) Free volume distributions for monodisperse spheres with selected numbers of contact neighbors, $q = 2, 4, 6, 8$, for $\phi_{init} = 0.4487$.}
\label{fig:fig1}
\end{figure}

\textcolor{black}{A {\it cavity} is a connected subset of the void
  space, which is defined as the volume that lies outside the union of
  spheres that define the system under consideration. For defining the
  free volumes, one considers {\it exclusion spheres} located at the
  coordinates of each hard sphere, with radius equal to the radius of
  the hard sphere {\it plus} the radius of the sphere whose free
  volume one wishes to compute. Given a configuration of $N$ hard
  spheres, the free volume of a given hard sphere is the volume of the
  {\it cavity} in which its coordinates lie, in a configuration of
  $N-1$ spherses excluding the sphere under consideration.  The
  algorithm for computing cavity volumes, described in detail in
  \cite{sastryalgo,maiti}, involves the following steps, for the
  calculation of free volumes in hard sphere configurations: (i) A
  Voronoi or generalized Voronoi (radical plane) construction. (ii)
  Identification of cavities based on connected Voronoi vertices that
  lie in the void space. (iii) Identification of the set of Delaunay
  tetraheda that enclose a given cavity. (iv) Computation of the
  cavity volume and surface area of the cavity contained in each
  Delaunay tetrahedron. An efficient way of implementing this
  algorithm to the free volume calculation without repeated
  tessellations is described in \cite{sastryfreevol,maiti}.}

The free volume distributions, over a wide range of densities\cite{sastryfreevol,maiti}, is described well by the form 
\begin{equation} 
f(v_f) \propto v_f^{\alpha} \exp(-\beta v_f^{\gamma})
\label{eq:fv} 
\end{equation} 
where the parameters $\alpha$, $\beta$ and $\gamma$ are smoothly
varying functions of density yielding a mean free volume that
decreases with density. Strictly at the jamming point, the free
volumes must equal zero, if one excludes 'rattlers',
  which are spheres with the number of contacts $q \le 3$, and are
  therefore free to move in some directions without hindrance from the
  contact neighbors. However, slightly below the jamming poing,
the free volumes are finite, and it is natural to ask what the free
volume distributions look like at such densities close to jamming. We
address this question in this paper, by generating jammed
configurations of monodisperse and bidisperse hard spheres,
\textcolor{black}{and calculating the free volume distributions of
  nearly jammed configurations obtained by decompressing them
  slightly.}


We generate jammed configurations following the procedure in
\cite{chaudhuri,berthier}, wherein starting from an equilibrated hard
sphere fluid at initial density $\phi_{init}$, a fast initial
compression is effected using a Monte Carlo simulation till the
pressure reaches $10^3$ (reduced units are used throughout). The
system is next treated as made of soft spheres \cite{ohern} (with
interaction potential $V_{\alpha \beta} = (1 -
\frac{r_{ij}}{\sigma_{\alpha \beta}})^2$ for $r_{ij} \le
\sigma_{\alpha \beta}$; = 0 for $r_{ij} > \sigma_{\alpha \beta}$,
where $\alpha$, $\beta$ represent different sizes of spheres $i$,
$j$), and compressed further, till the energy per particle is
$10^{-5}$. The system is then decompressed in steps, following energy
minimization at each step. A jammed configuration (and jamming density
$(\phi_J)$) is identified by the condition that the single particle
energy reaches $e = 10^{-16}$ or smaller. For data shown in figures 1
and 2, the jammed configurations are generated with a tolerance in
packing fraction of $10^{-5}$, and for figures 4 and 5, $10^{-8}$
\textcolor{black}{(based on the decompression steps used)}.
\textcolor{black}{We note that we consider only energy minimum
  structures throughout this procedure, and dynamical effects that may
  matter in real systems do not play a role in determining the jammed
  configurations.}

\begin{figure}[H]
\centering
\includegraphics[scale=0.3]{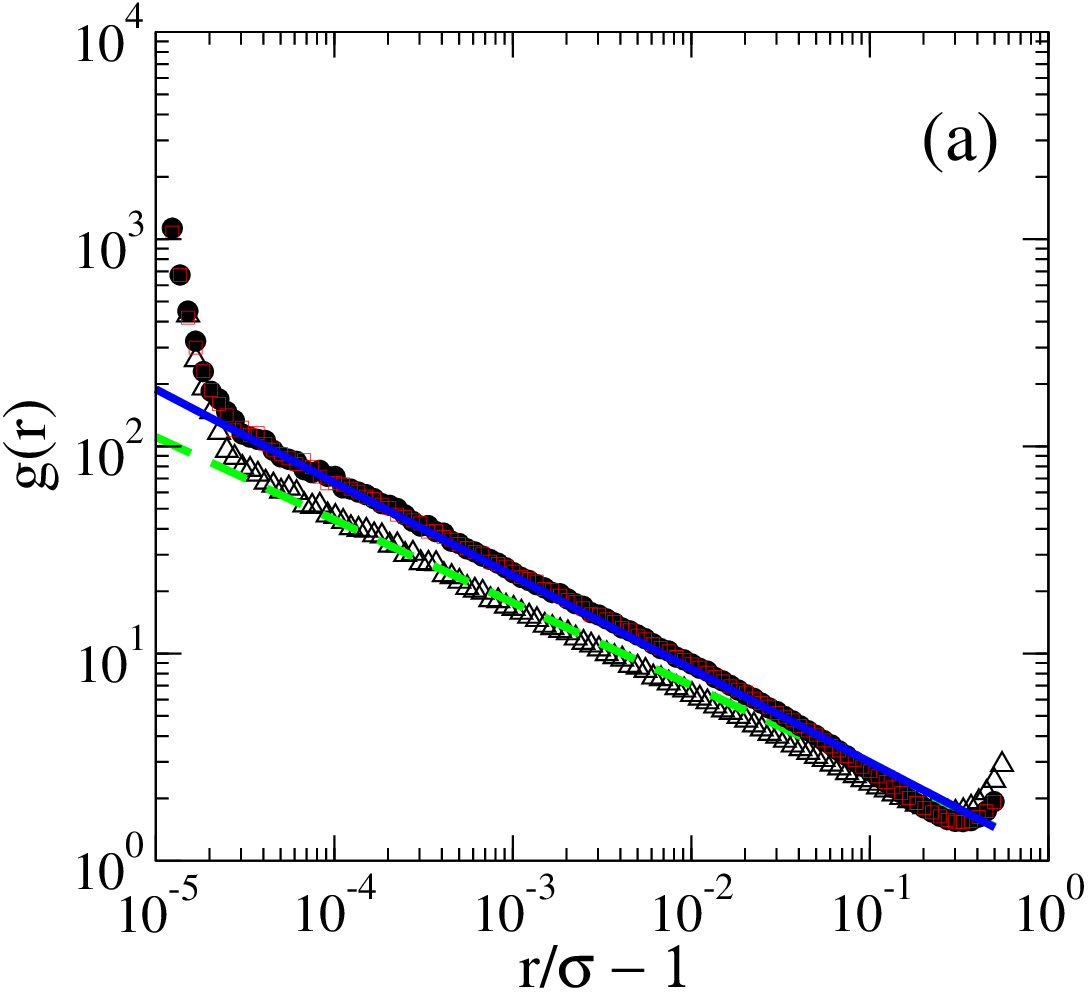}
\includegraphics[scale=0.3]{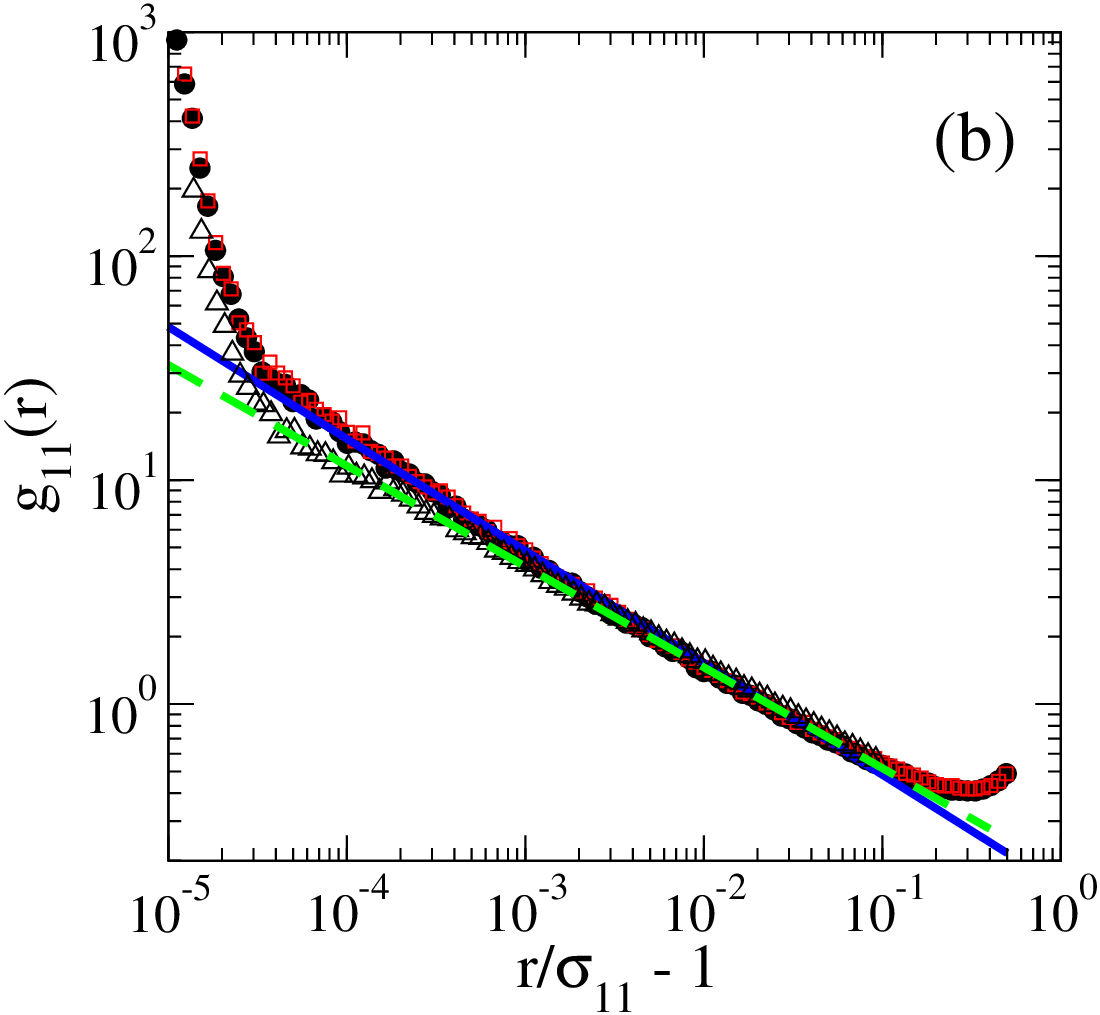}
\includegraphics[scale=0.3]{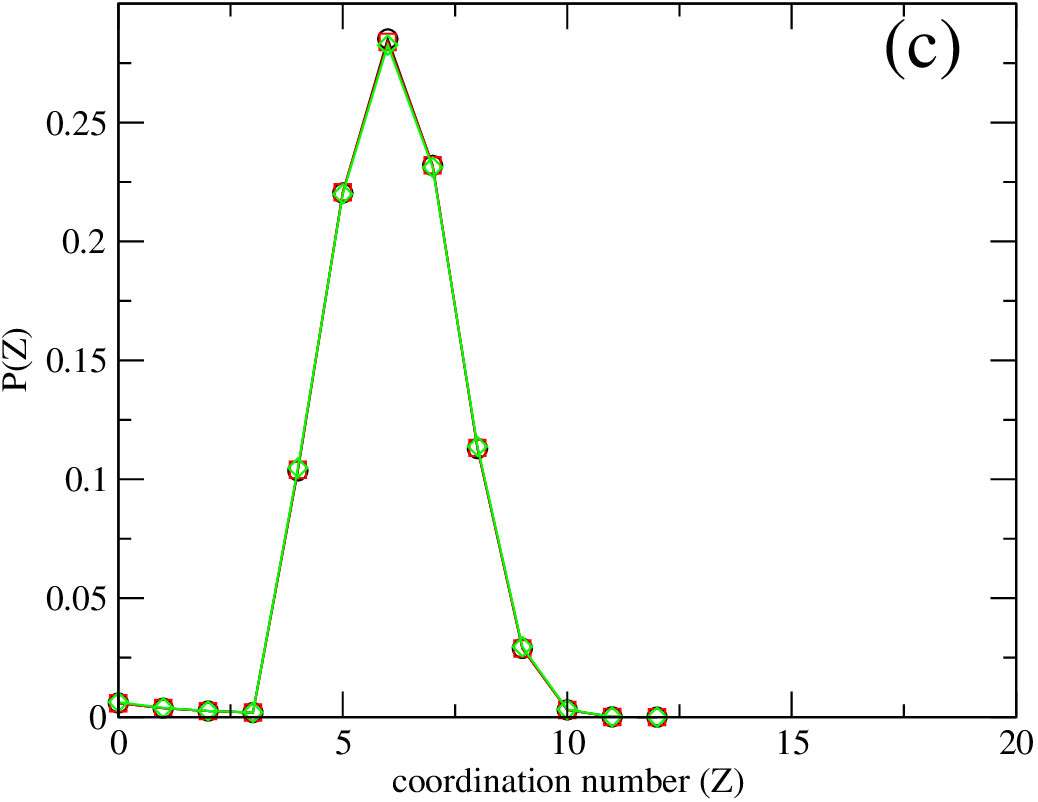}
\caption{Pair correlation function g(r) for (a) mono and (b) (component 1 of the) bidisperse sphere packings for the cases shown in Fig. 1. Power law fits close to contact are shown, with exponent $-0.45$ for monodisperse packings and $-0.5$ for the bidisperse packings. g(r) calculated excluding rattlers is also shown (open triangles) along with power law fit lines (dashed lines) for comparison. (c) The distribution of the number of contact neighbors for the monodisperse case.}
\label{fig:fig2}
\end{figure}

We study monodisperse and bidisperse hard sphere systems of size $N = 2000$. The bidisperse system is a 50:50 mixture of spheres of diameter $\sigma$ (component 1) and $1.4 \sigma$ (component 2). We choose three initial densities for the monodisperse fluid, which are smaller than the packing fraction ($0.545$) of spontaneous crystallization. Four different initial fluid densities of bidisperse hard spheres are chosen.  The jammed densities are evaluated after averaging over $500$ ($100$) independent configurations for figures 1,2 (figures 4,5), ensuring lack of crystallinity through a calculation of orientational order parameters. The statistics of jammed densities for different initial fluid densities are shown in the table I(monodisperse) and table II(bidisperse). 

\begin{table}[htbp]
   \centering
   \vspace{5mm}
   \begin{tabular}{| c | c | c |} \hline
          $\phi_{init}$   & $ \sim \phi_{J}(e\le 10^{-16})$ & Rattlers \\               \hline
           0.533           & 0.6419 & $1.6 \%$ \\ \hline
               0.4487     & 0.6394 & $ 1.7 \%$ \\ \hline
           0.2841        & 0.6392 & $1.7   \%$   \\ \hline
   \end{tabular}
   \caption{Statistics of jammed configurations of monodisperse spheres for system size $N = 2000$. $\phi_{init}$, $\phi_J$ are the initial and jammed configuration densities, and $e$ is the energy per particle. \textcolor{black}{Rattlers are
  spheres with number of contacts $q \le 3$.}}
   \label{tab:tab}
\end{table}

\begin{table}[htbp]
   \centering
   \begin{tabular}{| c | c | c | c |} \hline
          $\phi_{init}$   & $ \sim \phi_{J}(e\le 10^{-16})$ & Rattlers \\               \hline
           0.5919           & 0.6605 & $6.1 \% $\\ \hline
               0.5704     & 0.6554 & $ 5.6  \% $ \\ \hline
           0.5396        & 0.6498 & $4.9    \% $   \\ \hline
           0.363          & 0.6476 & $4.6   \% $  \\ \hline
   \end{tabular}
   \caption{Statistics of jammed configurations of bidisperse spheres for system size $N = 2000$. Symbols as in table I.}
   \label{tab:tab1}
\end{table}

We calculate the distribution of free volumes, using the algorithm
described in \cite{sastryfreevol,maiti}, for configurations that are
decompressed (by $10^{-8}$ in packing fraction) from the jammed
configurations and \textcolor{black}{thermalized by performing
 a short Monte Carlo simulation of $1000$ attempted random
  displacements (of magnitude comparable to the distance created
  between contact neigbhors by the decompression) per sphere.}
By extrapolating the parameters of Eq. \ref{eq:fv}, one may
expect a very narrow free volume distribution near the jamming point.
Fig.~\ref{fig:fig1} shows the free volume distribution near jamming
for both the monodisperse and bidisperse packings. In addition to a
peak at small free volumes, the distribution displays an unexpected
power law tail $(\approx v_f^{-\textcolor{black}{\delta}})$, over eight
orders of magnitude of free volumes. For both mono- and (both
components of the) bidisperse packings, the exponent
$\textcolor{black}{\delta} = -0.9$. These distributions, which are very
different from the extrapolation from the fluid state, do not depend
on the jamming density. We verify that the tail is not due to
rattlers. As shown in Fig.~\ref{fig:fig1} (a), eliminating rattlers
from consideration alters the exponent in the power law but does not
eliminate the feature. We also verify that the Lubachevsky-Stillinger
procedure\cite{lubachevsky} for generating jammed configurations also
produces the power law tail. As supported by results discussed below,
the observed power law tail is thus a structural signature of nearly
jammed particle packings.

Since the free volume of a sphere is determined by the location of its
neighbors, the power law tail in the free volume distribution may be
related to the pair correlation function $g(r)$, which, in addition to
a delta function at contact, has a well-studied singularity near
contact, $g(r) \approx (\frac{r}{\sigma} -
1)^{-\textcolor{black}{\eta}}$
\cite{silbert,donev,chaudhuri,charbonneau,wyart} for jammed
configurations. The connection is not straight-forward, since $g(r)$
is a two-body property, whereas the free volume of a sphere is a
many-body property that depends on the positions of all its
neighbors. Further, it is reasonable to expect, as indeed generally
done, that the contact neighbors bound the (compact) free volume of a
sphere in slightly decompressed configurations. The possibility that
non-contact neighbors may play a role at all is therefore suggested by
our surprising finding of a large free volume tail, which we explore.
As a preliminary test, we compute the free volume distribution for
groups of spheres each of which has a particualr number of contact
neighbors (we discuss the distribution of contact numbers further
below). These partial distributions are shown in Fig. 1(c), for
contact number $q = 2, 4, 6, 8$. While for $q = 6, 8$ the partial
distributions are confined to be small free volume peak, for $q = 2,
4$, one has a substantial contribution in the tail. Thus, the number
of contact neighbors plays a crucial role in determining the free
volumes, and for contact neighbors less than $6$, the free volumes are
significantly distributed in the tail. 

In Fig.~\ref{fig:fig2} (a) and (b) we show the $g(r)$, which exhibits
a power law regime near contact with exponent value
$\textcolor{black}{\eta} = -0.45$ for monodisperse packings and
$\textcolor{black}{\eta} = -0.5$ for component 1 of the bidisperse
case. As discussed previously \cite{silbert,donev,charbonneau}, and we
show in Fig.~\ref{fig:fig2}, the $g(r)$ exponent depends whether
rattlers are included in its evaluation or not.  We analyze
configurations which include rattlers, and choose the distance at
which $g(r)$ deviates from the power law, $(\approx 10^{-5})$, as the
cutoff to identify contact neighbors. The distribution of the number
of contact neighbors is shown in Fig. ~\ref{fig:fig2}(c) for the
monodisperse case.  Interestingly, roughly a third of the spheres have
less than $6$ contact neighbors, and thus the spheres whose free
volumes may contribute to the tail of the distribution is not a
negligible fraction. \textcolor{black}{We note that the population with
  $q \le 3$ ({\it i. e.} rattlers) are expected to have $q = 0$ with a
  precise definition of contact neighbors, and indeed find that the
  population of spheres with $q = 1,2,3$ decreases with an increase in
  the precision with which contact neighbors are identified.}

\begin{figure}[H]
\centering
\includegraphics[scale=0.3]{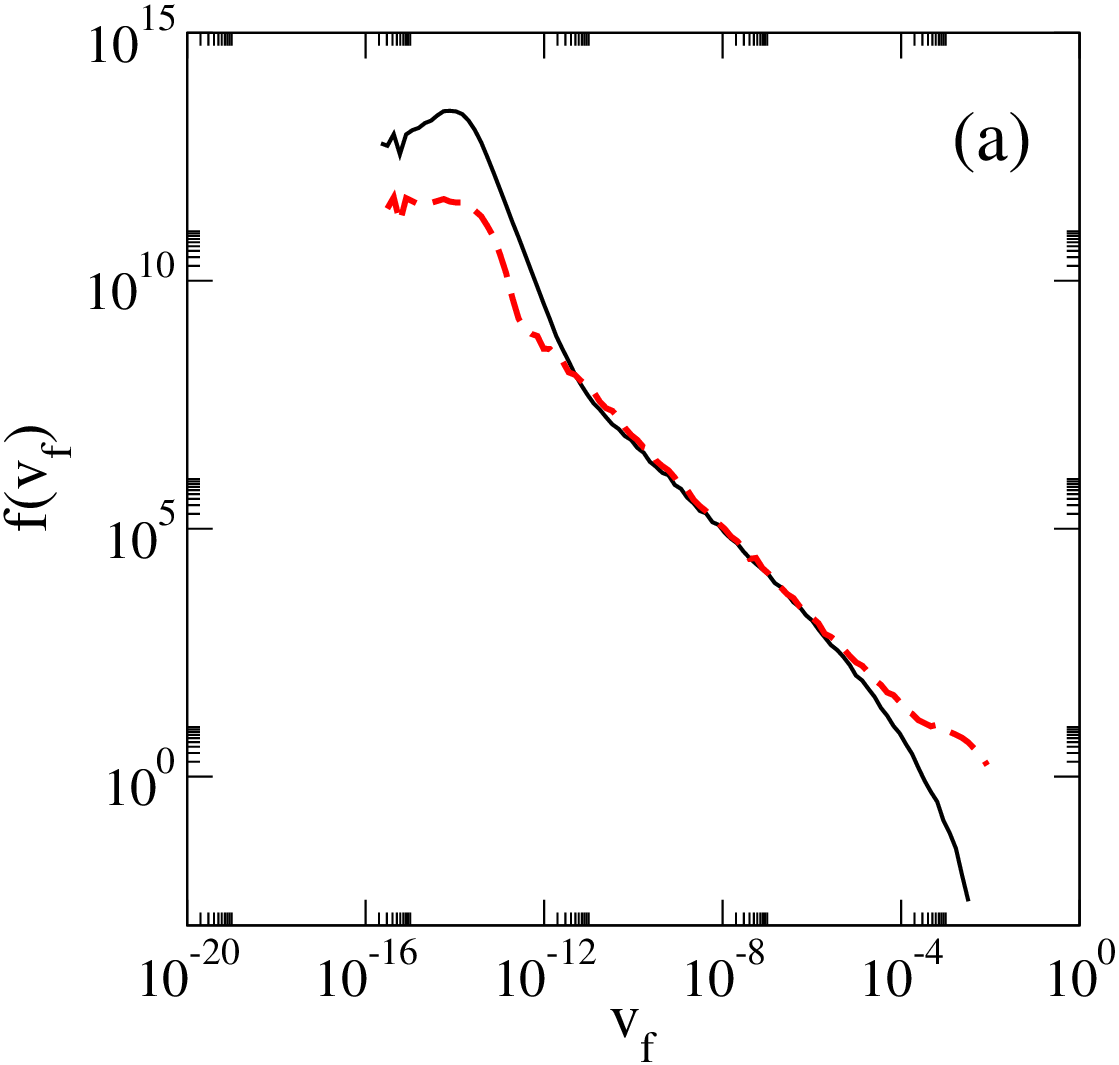}
\includegraphics[scale=0.3]{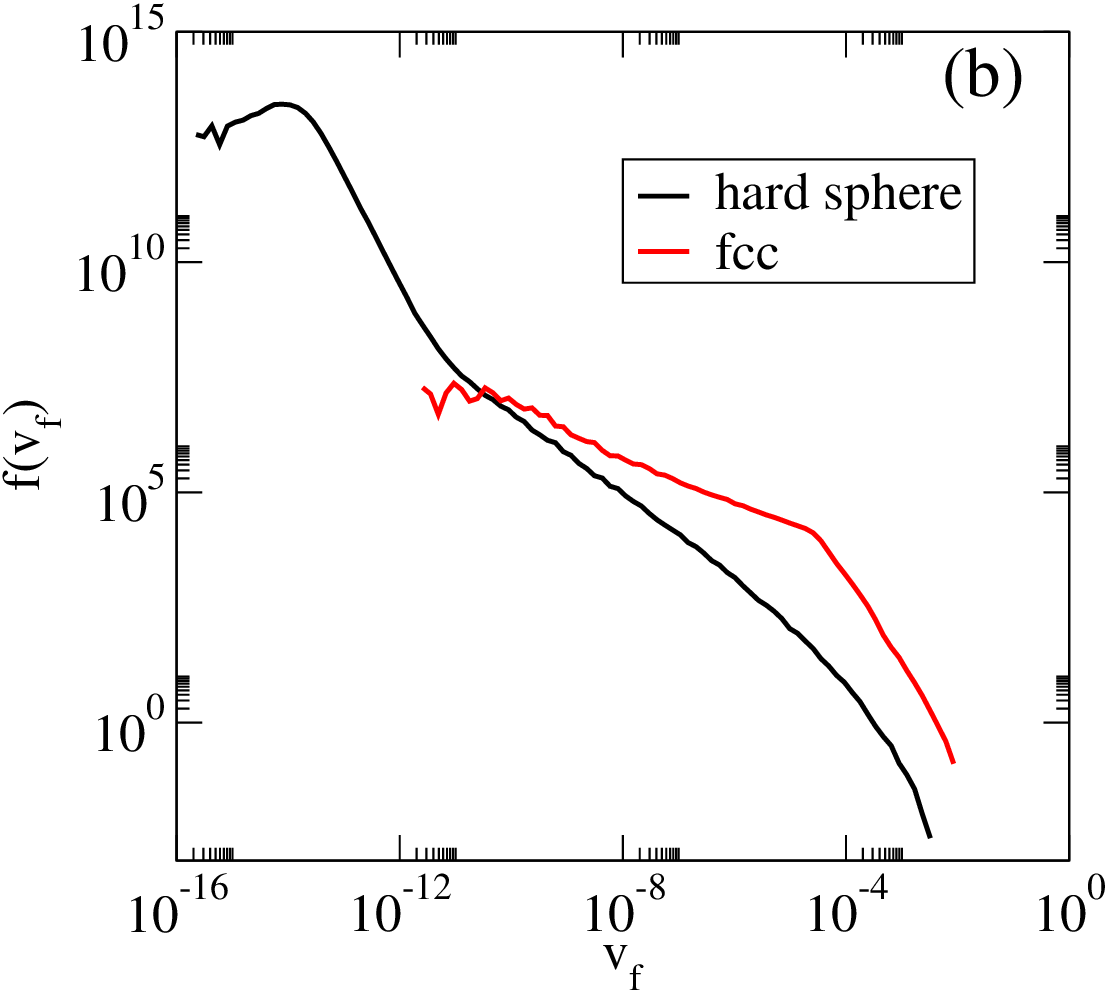}
\caption{(a) Comparison of the free volume distribution of monodisperse sphere packings (solid line) with the {\it icosahedron model}. (dashed line). (b) Comparision of the free volume distribution of monodisperse sphere packings (solid line) with the model calculation wherein the neighbors occupy nearest neighbor positions in the FCC lattice.}
\label{fig:fig3}
\end{figure}

We next consider a {\it toy} computation wherein the
  free volume of a sphere in the centre of an icosahedral cluster of
  spheres is considered. Starting with a regular, compact
  configuration, the $12$ spheres on the periphery are displaced
  according to the distribution of neighbor distances observed for
  monodisperse packings. As shown in Fig.~\ref{fig:fig3} (a), the free
  volume distribution for the {\it icosahedron model} indeed displays
  a power law tail, with the correct exponent, though it is too simple
  a model to capture all relevant details accurately, such as the
  relative amplitudes of the peak and the power law tail. Choosing
  initial neighbor positions arranged in an $FCC$ configuration,
  rather than an icosahedron, leads to a distribution with different
  features and a different exponent for its tail, as shown in Fig. Fig.~\ref{fig:fig3} (b) These observations,
  suggest that the free volume distribution depends sensitively on the
  many body correlations of neighbor positions, which needs further
  analysis to elucidate.

\begin{figure}[H]
\centering
\includegraphics[scale=0.3]{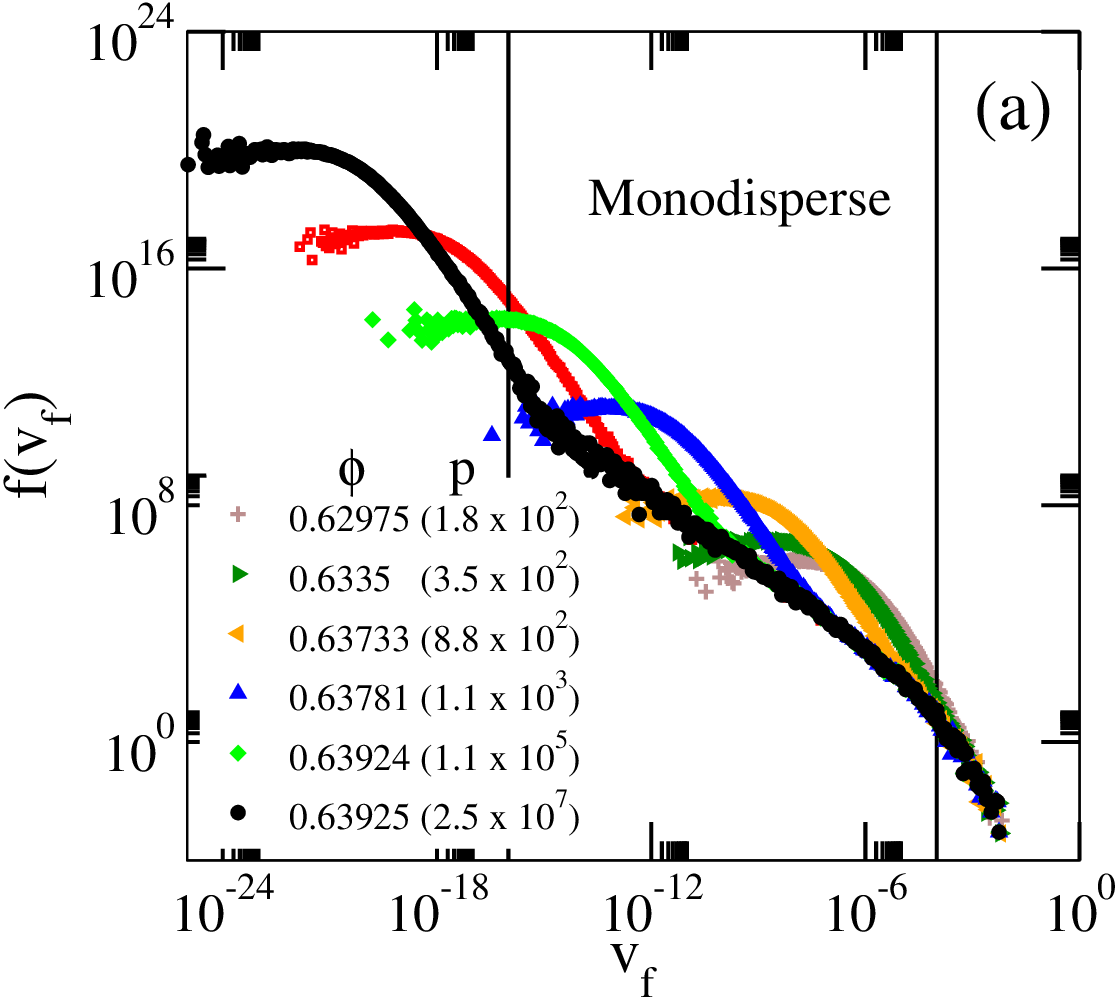}
\includegraphics[scale=0.3]{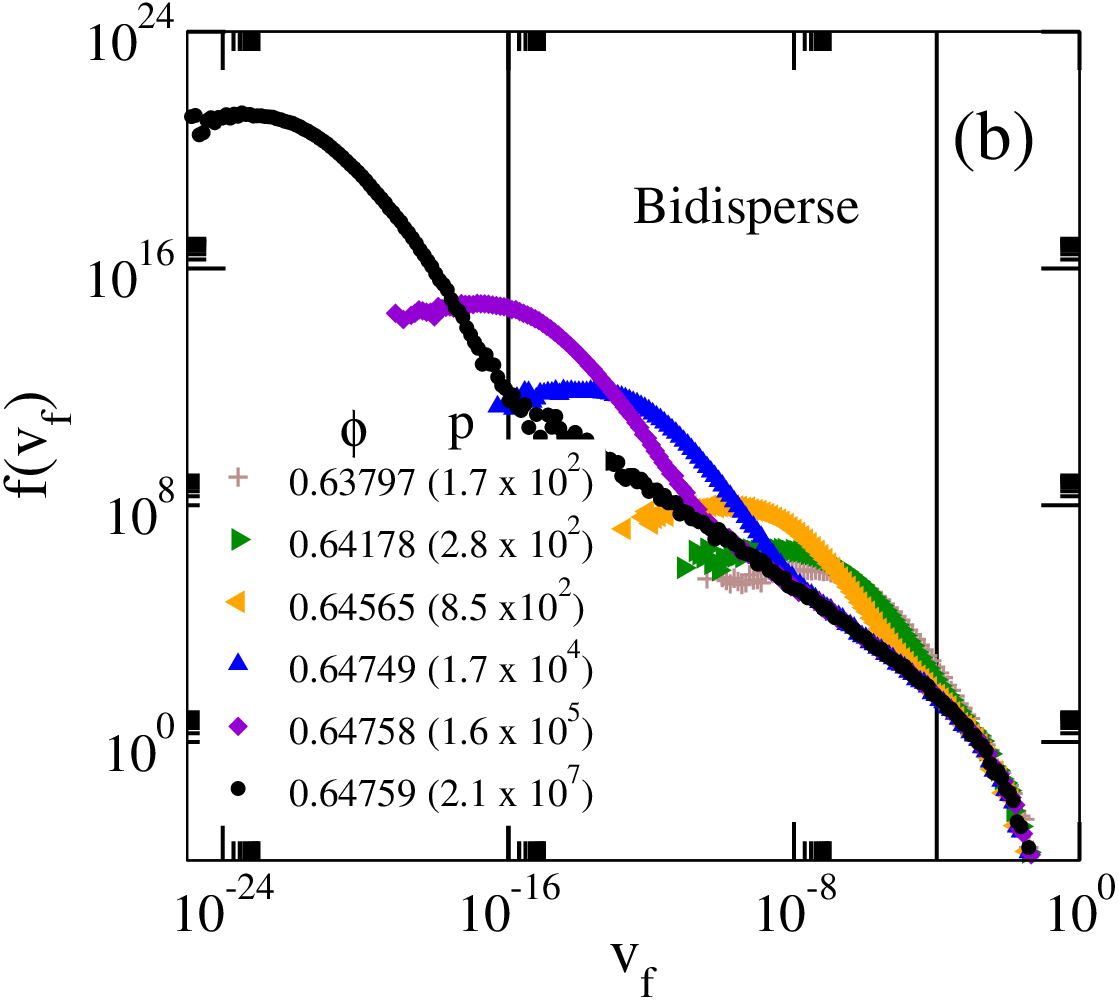}
\includegraphics[scale=0.3]{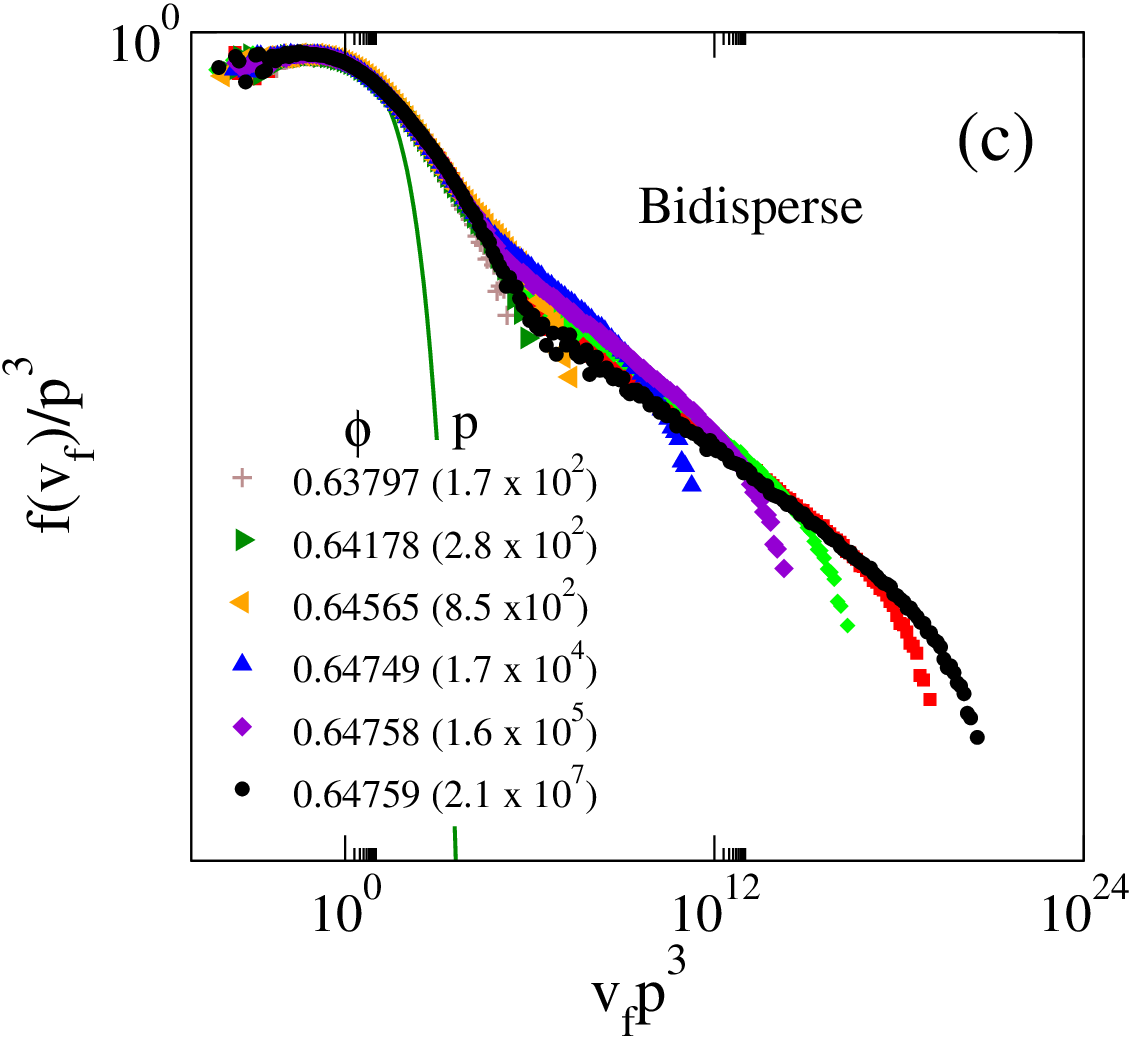}
\caption{Evolution of the free volume distribution of (a) monodisperse and (b) (component 1 of the) bidisperse sphere packings, upon decompression. 
Distributions for the bidisperse case have been scaled with $p^3$ in (c) 
($p = {P \over \rho k_B T} - 1$) to emphasize that invariant shape of the peak of the distributions, which is significantly broader than for the liquid (Eq. \ref{eq:fv} and Fig. \ref{fig:fig1}; solid line). The densities $\phi$ and  $p$ of (some of) the decompressed configurations are shown in the legends. Vertical lines in panel (a) indicate the range of $v_f$ over which one observes the power law tail for the highest density in both the cases.}

\label{fig:fig4}
\end{figure}

We next consider the manner in which the free volume distribution
evolves upon decompression. To do so, the sizes of the spheres are
rescaled to correspond to a range of densties lower than the jamming
density, and the configutations are thermalized by performing
\textcolor{black}{short Monte Carlo simulations (see above) at each
  density.} As shown in Fig. \ref{fig:fig4} (a) and (b), the distributions evolve
towards those found in the liquid state, but interestingly, the power
law tail persists over a finite range of densities.

\begin{figure}[H]
\centering
\includegraphics[scale=0.3]{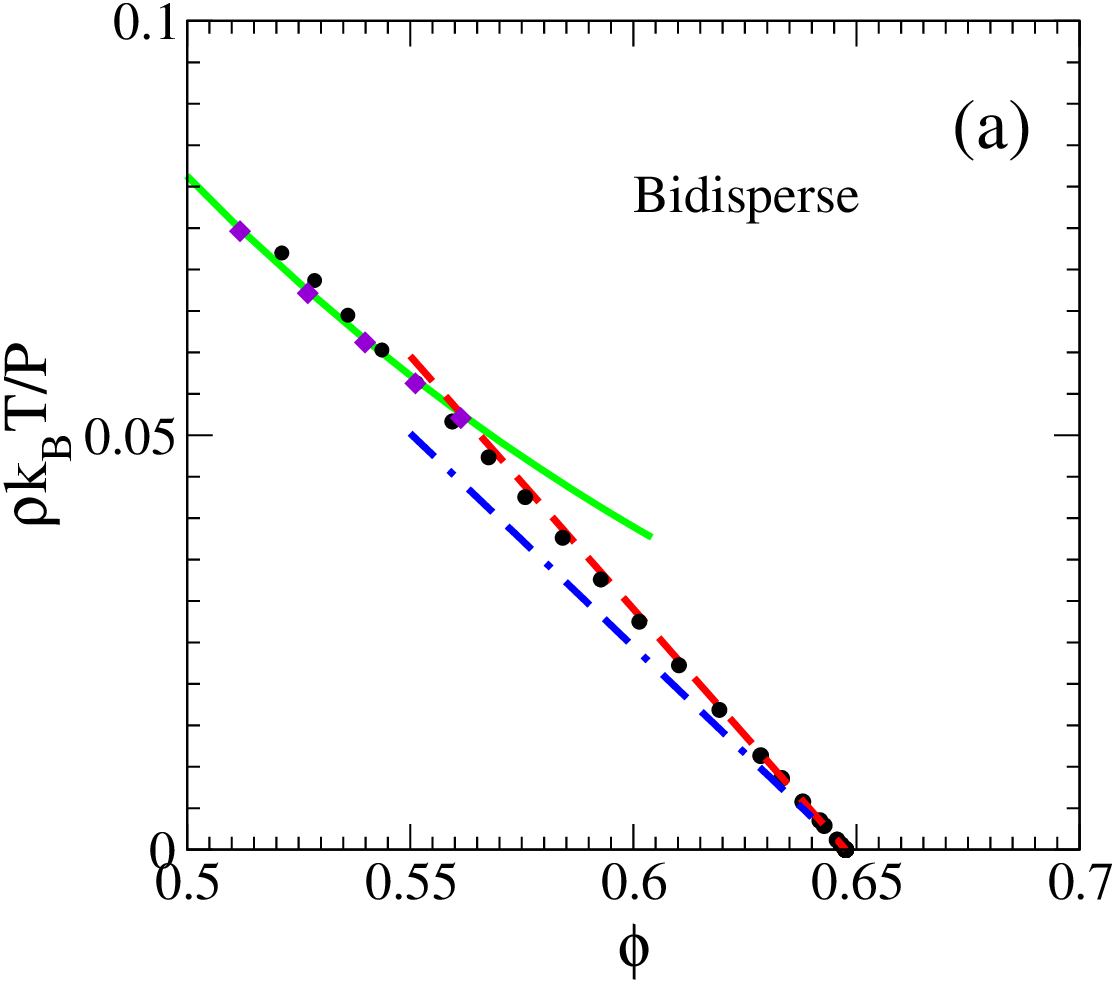}
\includegraphics[scale=0.3]{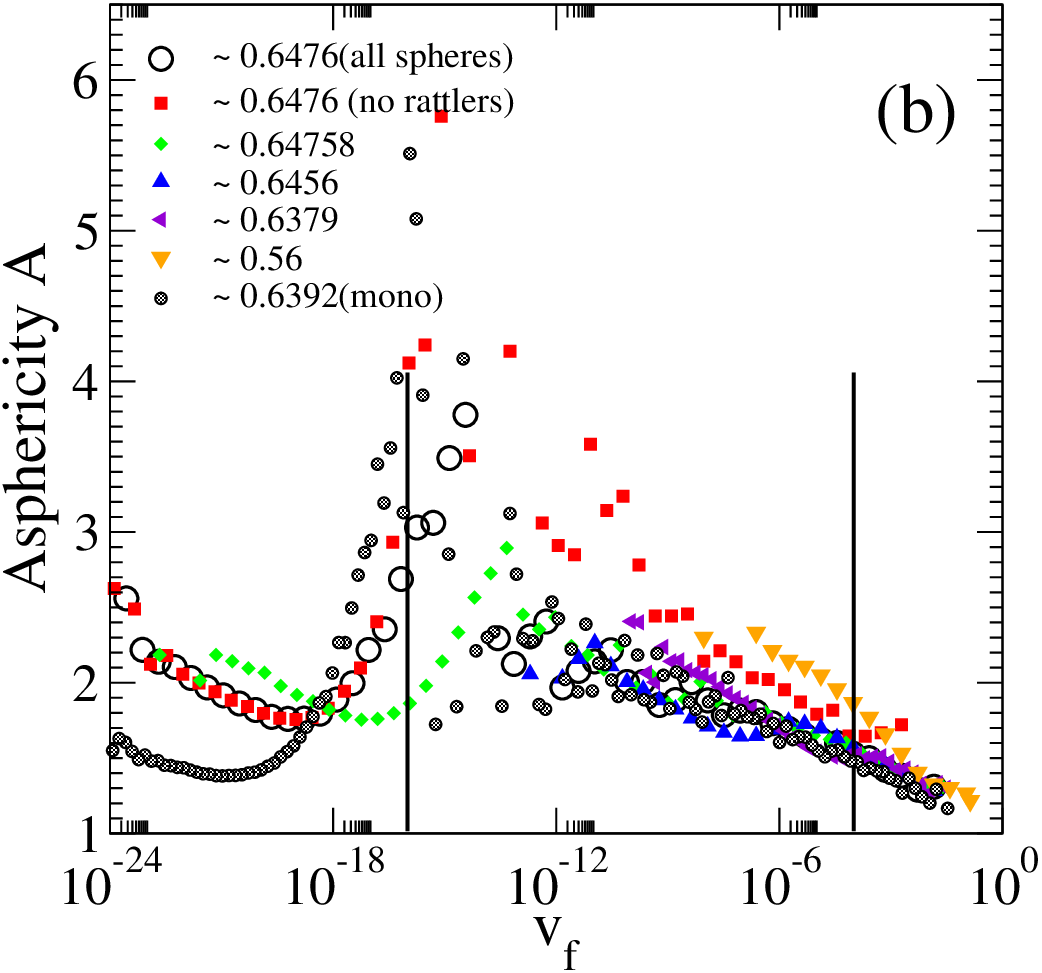}
\caption{(a) The equation of state (EOS) near jamming of bidisperse system. The solid line represents the Carnahan-Starling (CS) EOS. \textcolor{black}{Also shown (diamonds) are pressures for the fluid obtained from Eq. \ref{eq:fveos}.}  The data points from decompressed configurations are fit to the form \ref{eq:eosjp} with $C = 2.53$ (dashed lines) with an intersection with the CS EOS at $\phi = 0.562$. The line with $C = D$ is also shown for reference (dot dashed lines). (b) \textcolor{black}{Asphericity $A$ (see text) shown for bidisperse configurations for a range of densities. Aspericities for the monodisperse case, and the fluid ($\phi = 0.56$) are shown for comparison.}}
\label{fig:fig5}
\end{figure}

We use the configurations generated above to calculate the pressures, using Eq. \ref{eq:fveos}. The EOS near the jamming point predicted by Salsburg and Wood \cite{wood}, \textcolor{black}{obtained by assuming compact free volumes}, 
\begin{equation}
\frac{P}{\rho k_{B}T} =   \frac{C \phi_{J}}{\phi_{J} - \phi}\mbox{  ; $C = D$}.
\label{eq:eosjp} 
\end{equation}

This prediction has been verified \cite{donev} in earlier work but for
configurations without rattlers. Speedy \cite{speedy1} reported
$C=2.67$ for monodisperse hard spheres. As discussed in
\cite{donev,charbonneau_2}, the presence of rattlers will reduce the
value of $C$ linearly with the fraction of rattlers. However, our
results for the bidisperse system, shown in Fig. \ref{fig:fig5}(a),
consistent with a value of \textcolor{black}{ 
$C = 2.53$, cannot be explained by the presence of rattlers alone (see
  Table II, $\phi_{init} = 0.363$. Rattler fraction $f = 4.6\%$
  implies $C = D(1-f) = 2.862$), and must also have a contribution from
  free volume heterogeneity\cite{donev,charbonneau_2}. As can be seen
  from Fig. \ref{fig:fig4}(c), the peak of the free volume
  distribution, plotted against free volume scaled by $p^3$ (based on
  $p \sim 1/\delta$, $v_f \sim \delta^3$ where $\delta$ is the
  distance to the contact neighbors) has a density invariant shape
  that is significantly broader than the form obeyed by the liquid. We
  also calculate the asphericity $A$ of free volumes, defined as $A =
  {1 \over 2D} {s_f \over v_f^{2/3}}$ ($A = 0.806$ for spherical and $
  = 1$ for a cubic free volumes). As shown in Fig. \ref{fig:fig5}(b), free
volumes over the full range of values show asphericity to varying
degrees, for different degrees of decompression, with the
asphericities being maximum in the initial part of the power law tail
in all cases. The role played by these two factors merits further investigation.}
It is
interesting to note that the intersection of the free volume EOS
Eq. \ref{eq:eosjp} with the fluid (CS) EOS occurs around $0.562$, close to
the experimental number of $0.58$ for the glass transition as
described by mode coupling theory. Since the EOS Eq. \ref{eq:eosjp} is
obtained for jammed configurations decompressed with a very short
equilibration time, it is interesting to interpret the intersection
density of $0.562$ as the limiting density for the glass in the limit
of infinitely fast decompression, which has consistency with the
emergence of the mode coupling transition as a limit of stability in
some descriptions of the glass transition \cite{franz}. It should be
observed that the intersection density would be very much lower if $C
= D$. Thus, the presence of rattlers and free volume anisotropies seem
to have a role in determining a meaningful transition density from the
glass to the fluid states.

In summary, we have calculated free volume distributions for nearly jammed sphere packings, and show that they exhibit a characteristic power law tail, which we show is related to the power law singularity in the pair correlation function. The power law tail persists for moderate degrees of decompression and are thus a signature of packing close to jamming. We show evidence that the deviations of the equation of state close to jamming from the prediction for ideal jammed packings arise from the asphericities of the free volumes in addition to the presence of rattlers. 

\section{Acknowledgements:} We wish to thank Salvatore Torquato, Francesco Zamponi, Giorgio Parisi, Patrick Chabonneu, Gilles Tarjus and Sidney Nagel for useful discussions.

\eject 

\end{document}